\documentclass[preprint,aps,superscriptaddress,nofootinbib,tightenlines]{revtex4}
\usepackage{amsmath,amssymb}
\usepackage{graphicx}
\usepackage{bm}
\usepackage{comment}

\newcommand{\beq}{\begin{equation}}
\newcommand{\eeq}{\end{equation}}
\newcommand{\bea}{\begin{eqnarray}}
\newcommand{\eea}{\end{eqnarray}}

\def\OMIT#1{{}}

\newcount\hour \newcount\hourminute \newcount\minute 
\hour=\time \divide \hour by 60
\hourminute=\hour \multiply \hourminute by 60
\minute=\time \advance \minute by -\hourminute
\newcommand{\mydate}{\ \today \ - \number\hour :\number\minute}

\begin{document}
\begin{figure}[!t]
\vskip -1.5cm
\leftline{\includegraphics[width=0.25\textwidth]{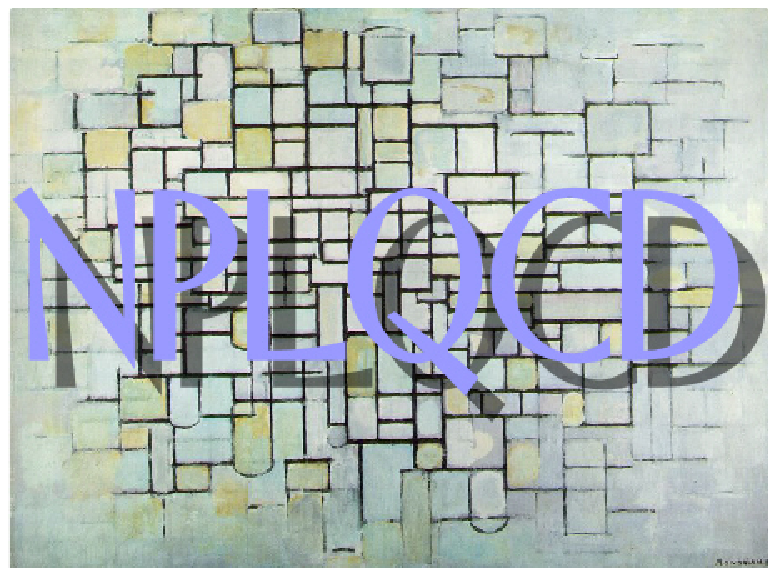}}
\end{figure}

\preprint{\vbox{
\hbox{UNH-07-03}
\hbox{UMD-40762-399}
\hbox{JLAB-THY-07-719}
\hbox{NT@UW-07-14}
}}

\vskip .5cm

\title{The $K^+ K^+$ Scattering Length from Lattice QCD}

\vskip .5cm
\author{Silas R.~Beane}
\affiliation{Department of Physics, University of New Hampshire,
Durham, NH 03824-3568.}
\author{Thomas C.~Luu}
\affiliation{N Division, Lawrence Livermore National Laboratory, Livermore, CA 94551.}
\author{Kostas Orginos}
\affiliation{Department of Physics, College of William and Mary, Williamsburg,
  VA 23187-8795.}
\affiliation{Jefferson Laboratory, 12000 Jefferson Avenue, 
Newport News, VA 23606.}
\author{Assumpta Parre\~no}
\affiliation{Departament d'Estructura i Constituents de la Mat\`{e}ria and
Institut de Ci\`encies del Cosmos, 
Universitat de Barcelona,  E--08028 Barcelona, Spain.}
\author{Martin J.~Savage}
\affiliation{Department of Physics, University of Washington, 
Seattle, WA 98195-1560.}
\author{Aaron Torok}
\affiliation{Department of Physics, University of New Hampshire,
Durham, NH 03824-3568.}
\author{Andr\'e Walker-Loud}
\affiliation{Department of Physics, University of Maryland, College Park, MD 20742-4111.}
\collaboration{ NPLQCD Collaboration }
\noaffiliation
\vphantom{}

\date{\mydate}

\vskip 0.8cm
\begin{abstract}
\noindent 
The $K^+ K^+$ scattering length is calculated in fully-dynamical
lattice QCD with domain-wall valence quarks on the MILC 
asqtad-improved gauge configurations with 
rooted staggered sea quarks.  
Three-flavor mixed-action chiral perturbation theory at next-to-leading order,
which includes the leading effects of the finite lattice spacing,
is used to extrapolate the results of the lattice calculation to the physical
value of $m_{K^+}/f_{K^+}$. 
We find $m_{K^+} \ a_{K^+ K^+} = -0.352 \pm 0.016$,
where the statistical and systematic errors have been combined in quadrature.
\end{abstract}
\maketitle



\vfill\eject

%
%
\section{Introduction \label{sec:Intro}}

\noindent
Strange hadrons may play a crucial role in the properties and evolution of 
nuclear material under extreme conditions~\cite{Page:2006ud}.
The interior of neutron stars provide one such environment in which the
densities are high enough that it may be energetically favorable
to have  strange baryons present in significant
quantities, depending upon their interactions with non-strange hadrons.
Further, it may be the case that a kaon condensate forms due to strong
interactions between kaons and nucleons~\cite{Kaplan:1986yq}.  
Unfortunately, the theoretical analysis of both scenarios is somewhat 
plagued by the limited knowledge of the interactions of
strange hadrons with themselves and with non-strange hadrons.

Heavy-ion collisions, such as those at the BNL Relativistic Heavy Ion
Collider (RHIC), also produce nuclear material in an extreme
condition.  Recent observations suggesting the formation of a
low-viscosity fluid are quite exciting as they provide a first glimpse
of matter not seen previously.  The late-time evolution of such a
collision requires an understanding of the interaction between many
species of hadrons, not just those of the initial state, including the
interactions between strange mesons and baryons.  While pion
interferometry in heavy-ion collisions is a well-established tool for
studying the collision region (for recent theoretical progress, 
see Refs.~\cite{Cramer:2004ih,Miller:2005ji,Miller:2007gh}), 
the STAR collaboration has recently
published the first observation of neutral kaon ($K_s^0$)
interferometry~\cite{Abelev:2006gu}.  In the analysis of
$K_s^0$-$K_s^0$ interferometry, the non-resonant contributions to the
final state interactions between the kaons were estimated using
three-flavor ($SU(3)_L\otimes SU(3)_R$) chiral perturbation theory
($\chi$-PT), the low-energy effective field theory of QCD.  Given the
sometimes poor convergence of $SU(3)_L\otimes SU(3)_R$ $\chi$-PT due
to the relatively large kaon mass compared to the scale of chiral
symmetry breaking ($\Lambda_\chi \sim 1$~GeV), particularly in the
baryon sector, it is important to be able to verify that the
non-resonant contributions to $KK$-scattering are indeed small, as
estimated in $\chi$-PT.

In this work we present the first lattice QCD calculation of
the $K^+ K^+$ scattering length. The calculations are performed on the coarse
MILC lattices (with a preliminary calculation on one ensemble of the fine MILC lattices) and
three-flavor mixed-action $\chi$-PT (MA$\chi$-PT), which includes the leading-order lattice-spacing effects,
is used to extrapolate to the continuum and to the (isospin-symmetric) physical value of the
meson masses. We find that at the physical value of $m_{K^+}/f_{K^+}$ 
\begin{equation}
        m_{K^+}\  a_{K^+ K^+} \ =\  -0.352 \pm 0.016
\ \ \  ,
\end{equation}
where the statistical and systematic errors have been combined in quadrature.

The $\pi$, $K$ and $\eta$ are identified as the pseudo-Goldstone
bosons associated with the spontaneous breaking of the approximate
chiral symmetry of quantum chromodynamics (QCD), and therefore the
form of their interactions is highly constrained.  In fact, at
leading order (LO) in $\chi$-PT, the scattering of two of these mesons
is uniquely determined~\cite{Weinberg:1966kf}.  
Corrections to the LO
scattering amplitude arise in a systematic expansion about the chiral
limit~\cite{Weinberg:1978kz,Gasser:1983yg}, 
scaling generically as $(m_{\pi,K,\eta}^2 / \Lambda_\chi^2)^n$ where
$n$ counts the order in the chiral expansion~\cite{Gasser:1984gg}.
For obvious reasons, $SU(3)$ $\chi$-PT is expected to converge more
slowly than two-flavor $\chi$-PT.

There is a wealth of phenomenological and theoretical knowledge
concerning low-energy $\pi\pi$ scattering.  The chiral extrapolation
formulae for $\pi\pi$ scattering are known to two loops, or
next-to-next-to-leading order (NNLO), in both
$SU(2)$~\cite{Bijnens:1995yn,Bijnens:1997vq} and
$SU(3)$~\cite{Bijnens:2004eu} $\chi$-PT.  Combined with a Roy equation
analysis~\cite{Roy:1971tc,Basdevant:1973ru,Ananthanarayan:2000ht},
this has allowed for remarkably-precise determinations of the two
$s$-wave $\pi\pi$ scattering
lengths~\cite{Colangelo:2001df,Caprini:2005an,Leutwyler:2006qq}.
In the case of
the $K\pi$ systems, the extrapolation formulae for the scattering
amplitudes are known to
one~\cite{Bernard:1990kw,Bernard:1990kx,Kubis:2001bx} and two
loops~\cite{Bijnens:2004bu} and have allowed for theoretical
predictions of the $I=1/2$ and $I=3/2$ scattering lengths.  However,
the uncertainty in these theoretical predictions is substantial.
There are proposed experiments to study the $K\pi$ atoms by the DIRAC
collaboration~\cite{DIRACprops} at CERN, J-PARC and GSI, which will
significantly reduce the uncertainty in these scattering lengths.
To date, there have been
no experimental determinations of the $I=1\ KK$ scattering length,
$a_{KK}^{I=1}$, but recently it has been calculated at next-to-leading order (NLO) in
$\chi$-PT~\cite{Chen:2006wf}.

The methods for studying two-particle interactions in a finite
Euclidean volume are well
known~\cite{Huang:1957im,Hamber:1983vu,Luscher:1986pf,Luscher:1990ux}.
The interaction energy of two hadrons in a finite volume uniquely
determines $p\cot \delta(p)$, and hence their scattering amplitude,
below kinematic thresholds.  The scattering parameters, such as the
scattering length and effective range, can then be determined from
calculations of $p\cot \delta(p)$ over a range of energies.  These
methods paved the way for pioneering quenched QCD calculations of
two-particle interactions a little over a decade
ago~\cite{Sharpe:1992pp,Gupta:1993rn,Kuramashi:1993ka,Kuramashi:1993yu,Fukugita:1994ve}.
Since then, there have been a number of additional quenched calculations
of the $I=2\ \pi\pi$ scattering
length~\cite{Fiebig:1999hs,Liu:2001zp,Liu:2001ss,Aoki:2002in,Aoki:2002ny,Juge:2003mr,Aoki:2005uf,Li:2007ey}.
The first dynamical calculation of $\pi\pi$ interactions (including
the phase-shift) was performed by the CP-PACS collaboration with two
flavors ($n_f=2$) of improved Wilson fermions~\cite{Yamazaki:2004qb}
and pion masses in the range $500 \lesssim m_\pi \lesssim 1100$~MeV.
Recently, dynamical calculations of the $I=2\ \pi\pi$ scattering
length with three flavors of light quarks ($n_f=2+1$) were performed
with pion masses in the range $300 \lesssim m_\pi \lesssim 500$~MeV.
These calculations used a mixed-action scheme of domain-wall valence
fermions on asqtad-improved staggered sea fermions at a single lattice
spacing of $b\sim 0.125$~fm~\cite{Beane:2005rj,Beane:2007xs}, and used
mixed-action $\chi$-PT (MA$\chi$-PT) (which describes the
finite-lattice spacing effects) calculations of the scattering
length~\cite{Chen:2005ab} to extrapolate to the physical meson masses.
Only recently has the first calculation of the $I=3/2\ K\pi$
scattering length in quenched QCD been performed~\cite{Miao:2004gy},
and a fully-dynamical $n_f=2+1$ calculation~\cite{Beane:2006gj}
followed shortly afterward.  When combined with $\chi$-PT, this latter
calculation allowed for a simultaneous prediction of the $I=1/2$ and $I=3/2$ $\ K\pi$
scattering lengths using the NLO extrapolation
formulae~\cite{Beane:2006gj}.

This paper is organized as follows.  Section~\ref{sec:Method} contains the
details of our mixed-action lattice QCD calculation.
Discussion of the relevant correlation functions and an outline of the
methodology and fitting procedures can also be found in this section.
The results of the lattice calculation and the analysis with
MA$\chi$-PT are presented in Section~\ref{sec:Extrapolate}. In this
section, the various sources of systematic uncertainty are identified
and quantified.  In Section~\ref{sec:Conclude} we conclude.

%
%
\section{Methodology and Details of the Lattice Calculation \label{sec:Method}}

\noindent
In calculating the $K^+ K^+$ scattering length,
the mixed-action lattice QCD scheme developed by the LHP
Collaboration~\cite{Renner:2004ck,Edwards:2005kw} was used
in which domain-wall
quark~\cite{Kaplan:1992bt,Shamir:1992im,Shamir:1993zy,Shamir:1998ww,Furman:1994ky}
propagators are generated from a smeared source on $n_f = 2+1$
asqtad-improved~\cite{Orginos:1999cr,Orginos:1998ue} rooted staggered
sea quarks~\cite{Bernard:2001av}.  
To improve the chiral symmetry
properties of the domain-wall quarks, hypercubic-smearing
(HYP-smearing)~\cite{Hasenfratz:2001hp,DeGrand:2002vu,DeGrand:2003in}
was used in the gauge links of the valence-quark action.  
In the sea-quark sector, there has
been significant debate regarding the validity of taking the fourth
root of the staggered fermion determinant at finite lattice
spacing~\cite{Durr:2004as,Durr:2004ta,Creutz:2006ys,Bernard:2006zw,Bernard:2006vv,Creutz:2007nv,Bernard:2006ee,Bernard:2006qt,Creutz:2007yg,Creutz:2007pr,Durr:2006ze,Hasenfratz:2006nw,Shamir:2006nj,Sharpe:2006re}.
While there is no proof, there are arguments to
suggest that taking the fourth root of the fermion determinant
recovers the contribution from a single Dirac 
fermion~\footnote{For a
nice introduction to staggered fermions and the fourth-root trick, see
Ref.~\cite{degrandANDdetar}.}.
The results of
this paper assume that the fourth-root trick recovers the correct
continuum limit of QCD.

The present calculations were performed predominantly with the coarse MILC
lattices with a lattice spacing of $b\sim 0.125$~fm, and a spatial
extent of $L\sim 2.5$~fm.  On these configurations, the strange quark
was held fixed near its physical value while the degenerate light
quarks were varied over a range of masses; see
Table~\ref{tab:MILCcnfs} and
Refs.~\cite{Beane:2006mx,Beane:2006pt,Beane:2006fk,Beane:2006kx,Beane:2006gf}
for details.  
Further, preliminary calculations were performed on 506 configurations of
one fine MILC ensemble. 
On the coarse MILC lattices, Dirichlet
boundary conditions were implemented to reduce the original time
extent of 64 down to 32 and thus save a factor of two in computational time. 
While
this procedure leads to minimal degradation of a nucleon signal, it
does limit the number of time slices available for fitting meson
properties. By contrast, on the fine MILC ensemble, anti-periodic
boundary conditions were implemented and all time slices are available.
%
%
\begin{table}[t]
 \caption{The parameters of the MILC gauge configurations and
   domain-wall propagators used in this work. The subscript $l$
   denotes light quark (up and down), and  $s$ denotes the strange
   quark. The superscript $dwf$ denotes the bare-quark mass for the
   domain-wall fermion propagator calculation. The last column is the 
   number of configurations times the number of sources per
   configuration.}
\label{tab:MILCcnfs}
\begin{ruledtabular}
\begin{tabular}{ccccccc}
 Ensemble        
&  $b m_l$ &  $b m_s$ & $b m^{dwf}_l$ & $ b m^{dwf}_s $ & $10^3 \times b
m_{res}$~\protect\footnote{Computed by the LHP collaboration.} & \# of propagators   \\
\hline 
2064f21b676m007m050 &  0.007 & 0.050 & 0.0081 & 0.081  & 1.604 & 468\ $\times$\ 16 \\
2064f21b676m010m050 &  0.010 & 0.050 & 0.0138 & 0.081  & 1.552 & 658\ $\times$\ 20 \\
2064f21b679m020m050 &  0.020 & 0.050 & 0.0313 & 0.081  & 1.239 & 486\ $\times$\ 24 \\
2064f21b681m030m050 &  0.030 & 0.050 & 0.0478 & 0.081  & 0.982 & 564\ $\times$\ 8 \\
2896f2b709m0062m031 & 0.0062 & 0.031 & 0.0080 & 0.0423 & $\sim$ 0.25~\protect\footnote{Estimated on a small number of configurations.} & 506\ $\times$\ 1 \\
\end{tabular}
\end{ruledtabular}
\end{table}
To determine the light-quark masses, the domain-wall pion was tuned to
the lightest staggered pion to within a few
percent~\cite{Renner:2004ck,Edwards:2005kw}.  This choice is somewhat
arbitrary as the partially-quenched and mixed-action effective field
theories exist to describe this and other choices~\cite{Bar:2005tu}
(provided the meson masses remain in the chiral regime), with the
expression for the $I=1\ KK$ scattering length determined to NLO in
$\chi$-PT, PQ$\chi$-PT and MA$\chi$-PT in Ref.~\cite{Chen:2006wf}.%
\footnote{The PQ$\chi$-PT and MA$\chi$-PT expressions for the $I=1\
KK$ scattering length are identical in form at NLO.  This is not
unique to this quantity and can be understood on more general grounds,
as mixed-action theories with chirally-symmetric valence fermions
exhibit many universal features~\cite{Chen:2007ug}.}
The choice of tuning to the lightest taste of staggered meson mass, as
opposed to one of the other tastes, provides for the ``most chiral''
domain-wall mesons and therefore reduces the 
error in extrapolating to the physical point.  The mass splitting between
the domain-wall mesons and the staggered taste-identity mesons, which
characterizes the unitarity violations present in the calculation, is
then given by~\cite{Aubin:2004fs,Bernard:2006wx}
\begin{align}
        m_{\pi_I}^2 - m_{\pi_{dwf}}^2 \simeq b^2 \Delta_\textrm{I} &=
        0.0769(22) \textrm{ (l.u.)}
        \qquad \textrm{\it coarse}\, ;
        \nonumber\\
        &= 0.0295(27) \textrm{ (l.u.)}
        \qquad \textrm{\it fine}\, .
\end{align}

In order to determine the interaction energy between the two kaons,
both the single-kaon, $C_{K^+}(t)$, and two-kaon, $C_{K^+ K^+}(p,t)$,
correlation functions were computed, where $t$ is the Euclidean time separation
between the hadronic source and sink operators and $p$ denotes the
magnitude of the equal and opposite spatial momentum of each kaon.
The single-kaon correlation function is
\begin{equation}\label{eq:C_K}
        C_{K^+}(t) = \sum_\mathbf{x} \langle\, K^-(t,\mathbf{x})\ K^+(0,\mathbf{0})\, \rangle\, ,
\end{equation}
where the sum over all spatial sites projects onto the zero-momentum
state, $\mathbf{p}=0$.  A correlation function  which projects onto
the $K^+ K^+$ $s$-wave state in the continuum limit is
\begin{equation}\label{eq:C_KK}
        C_{K^+K^+}(p, t) = 
                \sum_{|{\bf p}|=p}\ \sum_{\bf x , y}
                        e^{i{\bf p}\cdot({\bf x}-{\bf y})} 
                        \langle\, K^-(t,{\bf x})\ K^-(t, {\bf y})\ K^+(0, {\bf 0})\ K^+(0, {\bf 0})\, \rangle\, .
\end{equation}
In eqs.~\eqref{eq:C_K} and \eqref{eq:C_KK}, $K^+(t,\mathbf{x}) =
\bar{s}(t,\mathbf{x}) \gamma_5 u(t,\mathbf{x})$ is a Gaussian-smeared
interpolating field for the charged kaon.  In the relatively-large
spatial volumes used in the calculation, the interaction energy between the two
kaons is a small fraction of the total energy, which is dominated by
the kaon masses.  To determine this energy, the ratio of
correlation functions,
\begin{equation}\label{eq:C_KK_C_K}
        G_{K^+ K^+}(p,t) \equiv \frac{C_{K^+K^+}(p,t)}{C_{K^+}(t) C_{K^+}(t)}
                \longrightarrow \sum_{n=0}^\infty \mathcal{A}_n e^{-\Delta E_n t}\, ,
\end{equation}
was constructed, with the arrow denoting the large-time,
infinite-number-of-gauge-configurations limit (far from the boundary).  
Due to the periodic boundary-conditions imposed on the propagators computed on
the fine lattices, the $K^+$ correlation function become a single $\cosh$
function far from the source, while the $K^+ K^+$ correlation function become
the sum of two $\cosh$'s, one depending upon $m_{K^+}$ and the other depending
upon $E_{K^+ K^+}$, leading to a non-trivial form for $G_{K^+ K^+}(p,t)$.
As an alternative method to calculating the
interaction energy (and a check of the systematics), a
Jackknife analysis of the difference between the energies extracted
from the long-time behavior of the double- and single-kaon correlation
functions individually was performed, finding results in agreement with those determined from
eq.~\eqref{eq:C_KK_C_K}.  The interaction energy is related to the
two-particle energy eigenvalues and twice the kaon mass,
\begin{equation}\label{eq:DeltaE}
        \Delta E_n \equiv E^{KK}_n - 2m_K
                = 2 \sqrt{p_n^2 +m_K^2} - 2 m_K\, .
\end{equation}
In the absence of interactions, the energy levels occur at values of
the momenta $\mathbf{p} = 2\pi \mathbf{j} / L$ where $\mathbf{j}$ is an integer-triplet, 
corresponding to the
allowed single-particle momentum modes in a cubic volume.  In the
interacting theory, the two-particle eigen-momenta, $p_n$, are shifted
from these values and can be determined from eq.~\eqref{eq:DeltaE} and
the calculated interaction energy.  The L\"{u}scher
formula~\cite{Huang:1957im,Hamber:1983vu,Luscher:1986pf,Luscher:1990ux}
can then be used to determine the infinite-volume scattering
parameters from the real part of the inverse scattering amplitude by
solving the equation
\begin{equation}\label{eq:pcotdelta_S}
        p \cot \delta(p) = \frac{1}{\pi L} \mathbf{S} \left( \frac{pL}{2\pi} \right)\, ,
\end{equation}
which is valid below the inelastic threshold.  The regulated three-dimensional sum is~\cite{Beane:2003da}
\begin{equation}\label{eq:S}
        \mathbf{S}(\eta) \equiv \sum_{\mathbf{j}}^{|\mathbf{j}| < \Lambda}
                \frac{1}{|\mathbf{j}|^2 - \eta^2} - 4\pi \Lambda\, ,
\end{equation}
which runs over all triplets of integers $\mathbf{j}$ such that
$|\mathbf{j}| < \Lambda$ and the limit $\Lambda \rightarrow \infty$ is
implicit.  The scattering parameters are then related to $p \cot
\delta(p)$ through the effective-range expansion
\begin{equation}
        p \cot \delta(p) = \frac{1}{a} + \frac{1}{2}r p^2 + \mathcal{O}(p^4)\, ,
\end{equation}
where $a$ is the scattering length and $r$ is the effective range.
For naturally-sized scattering lengths and small interaction momenta,
$p\cot \delta(p)$ is predominantly given by the inverse scattering
length.

%
%
\section{Analysis and the Chiral and Continuum Extrapolations \label{sec:Extrapolate}}

\begin{figure}[!t]
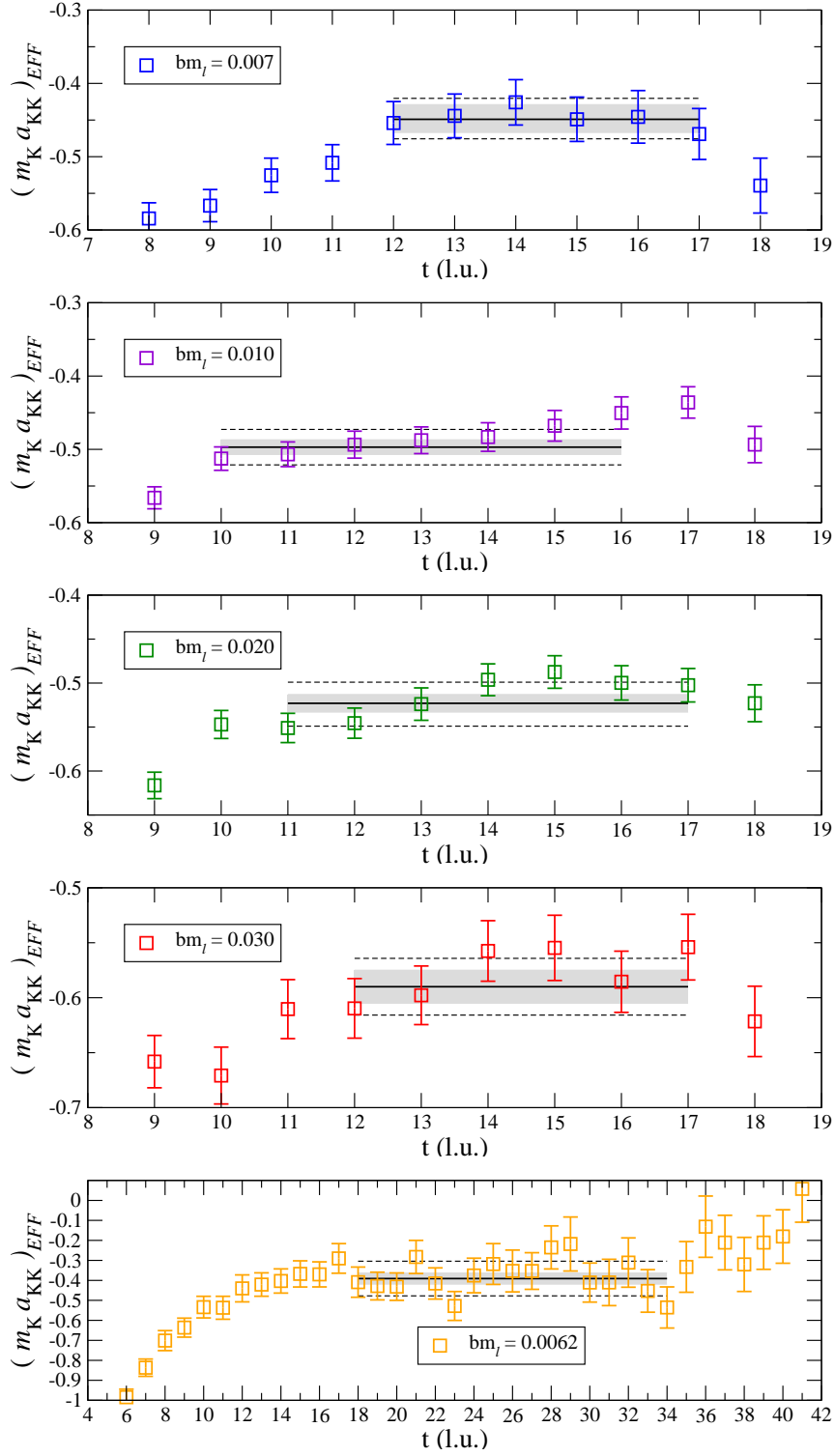

\center
\begin{tabular}{c}
\includegraphics*[width=0.7\textwidth,viewport=2 5 700 240,clip]{figures/EFFmKa007.eps}\\
\hfill
\includegraphics*[width=0.7\textwidth,viewport=2 5 700 240,clip]{figures/EFFmKa010.eps}\\
\hfill
\includegraphics*[width=0.7\textwidth,viewport=2 5 700 240,clip]{figures/EFFmKa020.eps}\\
\hfill
\includegraphics*[width=0.7\textwidth,viewport=2 5 700 240,clip]{figures/EFFmKa030.eps}\\
\hfill
\includegraphics*[width=0.7\textwidth,viewport=2 5 700 240,clip]{figures/EFFmKa0062.eps}\\
\end{tabular}
\caption{\label{fig:effective_a} \textit{ The effective $K^+K^+$
scattering length times the effective $m_{K^+}$ as a function of time
slice. The solid black lines and shaded
regions are fits with 1-$\sigma$ statistical uncertainties (Table~\ref{tab:latt_ma}).  
The dashed lines correspond to the statistical and systematic (Table~\ref{tab:latt_ma})
uncertainties added in quadrature.
}}
\end{figure}

\noindent
It is convenient to present the results of our calculation with
``effective scattering length" plots, determined from the ratio of
correlation functions,
\begin{equation}\label{eq:effene} 
        \Delta E_{K^+ K^+}(t) = \log\left({ G_{K^+ K^+}(0,t)\over  G_{K^+ K^+}(0,t+1)}\right)\, ,
\end{equation}
and similarly on the fine ensemble.  
For each time slice, $\Delta E(t)$ is inserted into
eq.~\eqref{eq:pcotdelta_S} which yields a 
scattering length at each time slice, $a_{K^+ K^+}(t)$.  
To remove any scale-setting ambiguities, the scattering length is multiplied by
the ``effective" kaon mass, $m_K(t)$.  
The effective scattering length plots associated with each lattice ensemble are
shown in Fig.~\ref{fig:effective_a}.
%
%
\begin{figure}[!t]
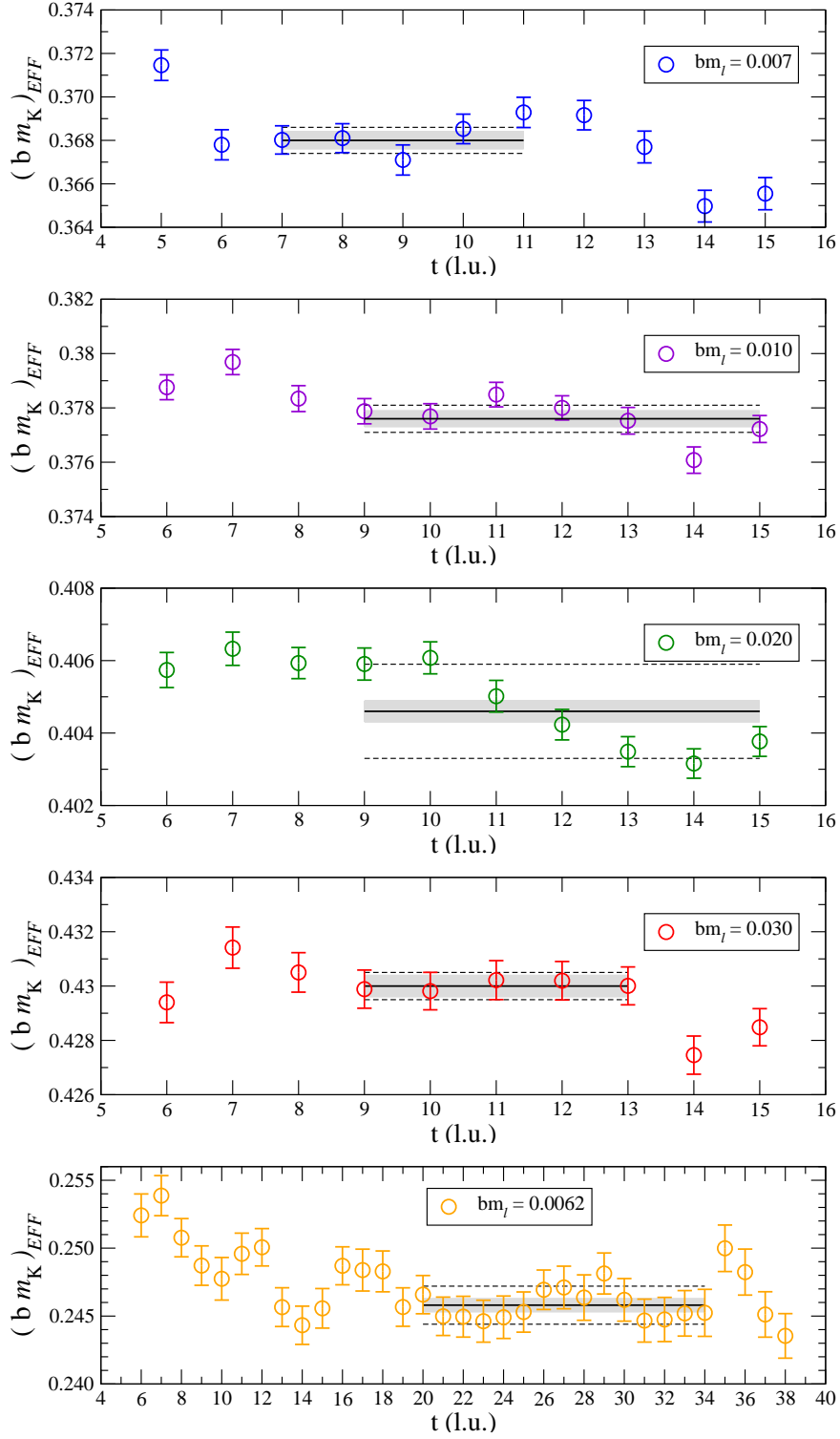

\center
\begin{tabular}{cc}
\includegraphics*[width=0.7\textwidth,viewport=2 5 700 240,clip]{figures/EFFkaon007.eps}\\
\hfill
\includegraphics*[width=0.7\textwidth,viewport=2 5 700 240,clip]{figures/EFFkaon010.eps}\\
\hfill
\includegraphics*[width=0.7\textwidth,viewport=2 5 700 240,clip]{figures/EFFkaon020.eps}\\
\hfill
\includegraphics*[width=0.7\textwidth,viewport=2 5 700 240,clip]{figures/EFFkaon030.eps}\\
\hfill
\includegraphics*[width=0.7\textwidth,viewport=2 5 700 240,clip]{figures/EFFkaon0062.eps}\\
\end{tabular}
\caption{\label{fig:effective_mK} \textit{ The effective $m_{K^+} (t)$
plots.  The solid black lines and shaded regions are fits with
1-$\sigma$ statistical uncertainties (Table~\ref{tab:latt_ma}).  
The dashed lines correspond to the statistical and systematic (Table~\ref{tab:latt_ma})
uncertainties added in quadrature.
}}
\end{figure}
The statistical errors are determined from a Jackknife analysis, while
the quoted systematic errors are estimated from both the range of fits as
well as the two methods of determining the interaction energy
described in Sec.~\ref{sec:Method}.  
In Table~\ref{tab:latt_ma} the calculated values of the meson masses, decay constants,
two-particle energy shifts and scattering lengths are presented. 
Effective kaon
mass plots and effective scattering length plots are shown in 
Fig.~\ref{fig:effective_mK} and Fig.~\ref{fig:effective_a},
respectively.
%
%

%
%
\begin{table}[t]
\caption{\label{tab:latt_ma} \textit{
Masses, energies and scattering lengths determined from the lattice calculation.
The first uncertainty assigned to each quantity is statistical, determined with the Jackknife
procedure, and  the second uncertainty is an estimated fitting systematic.}}
\begin{ruledtabular}
\begin{tabular}{c|cccc|c}
Quantity & $m_l = 0.007$ & $m_l = 0.010$ & $m_l = 0.020$ & $m_l = 0.030$ & $m_l = 0.0062$ \\ \hline
$b\ m_\pi$ & 0.1846(4)(2) & 0.2226(4)(3) & 0.3104(3)(15) & 0.3747(4)(8) & 0.1453(5)(13) \\
Fit Range & 8--14 & 9--13 & 9--15 & 6--13 & 17--39 \\
\hline
$b\ m_K$ & 0.3680(4)(4) & 0.3776(3)(4) & 0.4046(3)(13) & 0.4300(4)(3) & 0.2458(5)(13)\\
Fit Range & 7--11 & 9--15 & 9--15 & 9--13 & 20--34 \\
\hline
$ m_\pi / f_K$  & 1.712(4)(3) & 2.069(3)(5) & 2.835(3)(11) & 3.335(4)(9) & 1.978(15)(12) 
\\ 
$m_K / f_K$ & 3.412(5)(4) & 3.509(3)(6) & 3.695(3)(10) & 3.827(4)(9) & 3.344(19)(21) 
\\
\hline
$\Delta E_{KK}$(l.u.) 
& 0.00619(30)(32) & 0.00663(15)(35) & 0.00606(14)(22) & 0.00613(19)(10) & 0.00437(36)(105) \\
Fit Range & 12--17 & 10--16 & 11--17 & 12--17 & 18--34 \
\\
\hline
$m_{K^+} a_{K^+ K^+}$  
& -0.448(19)(20) & -0.497(10)(22) & -0.523(10)(23) & -0.590(15)(21) &
-0.391(28)(82) \\
$(b\neq 0)$ & & & & & \\
\end{tabular}
\end{ruledtabular}
\end{table}

%
%
\subsection{Mixed-Action $\chi$-PT at One Loop \label{sec:ResultsB}}

\noindent
The lattice QCD calculations performed in this work are isospin symmetric,
$m_u=m_d$, and do not include electromagnetism.
Therefore isospin is a good quantum number.
Having computed the $K^+ K^+$ scattering length at a number of unphysical pion
masses and at a finite lattice-spacing, isospin-symmetric MA$\chi$-PT is used to extrapolate
to the physical (isospin-symmetric) meson masses and to the continuum.

In Ref.~\cite{Chen:2006wf}, the expression for the $I=1\ KK$
scattering length was determined to NLO in $\chi$-PT, including
corrections due to mixed-action lattice artifacts.  As with the 
$I~=~2\ \pi\pi$ scattering length~\cite{Chen:2005ab}, it was demonstrated that
when the mixed-action extrapolation formula is expressed in terms of
the lattice-physical parameters computed on the lattice,~\footnote{
Quantities calculated directly from the correlation functions are
denoted as lattice-physical parameters.  These are not extrapolated to
the continuum, to infinite volume nor to the physical quark mass
point.}  $m_\pi$, $m_K$ and $f_K$, there are no
lattice-spacing-dependent counterterms at $\mathcal{O}(b^2)$,
$\mathcal{O}(b^2 m_K^2)$ or $\mathcal{O}(b^4)$.  There are finite
lattice-spacing-dependent corrections, proportional to $b^2
\Delta_\mathrm{I}$, and therefore entirely determined to this order in
MA$\chi$-PT.  Again, as with the $I=2\ \pi\pi$ system, the NLO MA
formula for $m_K a_{KK}^{I=1}$ does not depend upon the mixed
valence-sea meson masses, and therefore does not require knowledge of
the mixed-meson masses~\cite{Orginos:2007tw}.  This
allows for a precise determination of the predicted MA corrections to
the scattering length.  At NLO in MA$\chi$-PT, the scattering length
takes the form
\begin{multline}\label{eq:mKaKKMA}
m_K a_{KK}^{I=1}(b\neq 0) =-\frac{m_{K}^{2}}{8 \pi f_{K}^{2}} \bigg\{ 1
        +\frac{m_{K}^{2}}{ (4\pi f_{K})^{2}} \bigg[ 
                C_\pi \ln \left( \frac{m_\pi^{2}}{\mu^2} \right) 
                +C_K \ln \left( \frac{m_K^{2}\,}{\mu ^2} \right) 
        \\
                +C_X \ln \left( \frac{\tilde{m}_X^{2}}{\mu^2} \right) 
                +C_{ss} \ln \left( \frac{m_{ss}^{2}}{\mu ^{2}} \right) 
                + C_0 
        - 32(4\pi)^2\, L_{KK}^{I=1}(\mu)
        \bigg] \bigg\}\, ,
\end{multline}
where the various coefficients, $C_i$, along with $\tilde{m}_X^{2}$ and $m_{ss}^{2}$,
can be
found in Appendix E of Ref.~\cite{Chen:2006wf}.  
To account for the predicted MA corrections, one can either use eq.~\eqref{eq:mKaKKMA} to
directly fit the results of the lattice calculation (Table~\ref{tab:latt_ma}) or one
can determine the quantity
\begin{equation}
        \Delta_{MA} \left(m_K a_{KK}^{I=1} \right)
                = m_K a_{KK}^{I=1} \Big|_{MA} - m_K a_{KK}^{I=1} \Big|_{\chi PT}\, ,
\end{equation}
collected in Table~\ref{tab:mKaKK} and Table~\ref{tab:mKaKK_FINE}, 
subtract this from the results of the lattice calculation 
and use the NLO $\chi$-PT expression for the scattering length,
\begin{multline}\label{eq:mKaKKQCD}
m_K a_{KK}^{I=1} = -\frac{m_K^2}{8 \pi f_K^2} \bigg\{ 1 +\frac{m_K^2}{(4 \pi f_K)^2} \bigg[
        2 \ln \left( \frac{m_K^2}{\mu^2} \right)
        -\frac{2 m_\pi^2}{3 (m_\eta^2 -m_\pi^2)}\, \ln \left( \frac{m_\pi^2}{\mu^2} \right)
        \\
        +\frac{2(20m_K^2 -11m_\pi^2)}{27(m_\eta^2 -m_\pi^2)}\,
        \ln \left( \frac{m_\eta^2}{\mu^2} \right)
        -\frac{14}{9}
        -32(4\pi)^2\, L_{KK}^{I=1}(\mu)
        \bigg]
        \bigg\}\, .
\end{multline}
As there is only one counterterm at NLO, it can be determined on each ensemble.
In order to carry out this analysis, further
sources of systematic errors are identified; higher-order effects in the chiral
expansion, $\Delta_{NNLO}(m_K a_{KK}^{I=1})$; exponentially-suppressed
finite-volume effects, $\Delta_{FV}(m_K a_{KK}^{I=1})$; residual
chiral symmetry breaking effects from the domain-wall action,
$\Delta_{m_{res}}(m_K a_{KK}^{I=1})$; and the error in truncating the
effective-range expansion with the inverse scattering length,
$\Delta_{range}(m_K a_{KK}^{I=1})$.  These various sources
of systematic uncertainty, as well as the predicted mixed-action
corrections, the adjusted scattering lengths and the determined values
of $L_{KK}^{I=1}(\mu)$ are given in Table~\ref{tab:mKaKK}  and
Table~\ref{tab:mKaKK_FINE}.  
In the following
sections, each source of systematic uncertainty is addressed in
turn.
%
%
\begin{table}[t]
\caption{\label{tab:mKaKK} \textit{
The continuum limit of the scattering length at the physical point on the
coarse MILC lattices, the
extracted counterterm that enters at NLO in $\chi$-PT, and the various
systematic uncertainties that have been identified beyond those associated with
fitting.
The correction factors, $\Delta_i$,  are defined in the text.
The first uncertainty associated with each  scattering length is statistical,
the second is the systematic uncertainty from Table~\ref{tab:latt_ma} and the
third is from the systematic uncertainties presented in  this table (combined in
quadrature).  
The first uncertainty associated with each $L_{KK}^{I=1}(\mu=f_K)$ is 
statistical, while the second is systematic (all systematics combined in
quadrature).
}}
\begin{ruledtabular}
\begin{tabular}{c|cccc}
Quantity & $m_l = 0.007$ & $m_l = 0.010$ & $m_l = 0.020$ & $m_l = 0.030$ \\ \hline
$\Delta_{MA} \left(m_K a_{KK}^{I=1} \right)$ 
& -0.0067(14) & -0.0062(16) & -0.0052(19) & -0.0048(21) \\
$\Delta_{NNLO}\left(m_K a_{KK}^{I=1} \right)$  
& $\pm 0.016$ & $\pm 0.019$ & $\pm 0.028$ & $\pm 0.037$ \\
$\Delta_{FV} \left(m_K a_{KK}^{I=1} \right)$ 
& $\pm 0.001$ & $\pm 0.001$ & $\pm 0.000$ & $\pm 0.000$ \\
$\Delta_{m_{res}} \left(m_K a_{KK}^{I=1} \right)$ 
& $\pm 0.007$ & $\pm 0.006$ & $\pm 0.005$ & $\pm 0.004$ \\
$\Delta_{range}\left(m_K a_{KK}^{I=1} \right)$  
& $\pm $ 0.008 & $\pm 0.008$ & $\pm 0.008$ & $\pm 0.007$ \\
\hline
$m_{K^+} a_{K^+ K^+}$
& -0.441(19)(20)(19) & -0.491(10)(22)(22) & -0.518(10)(23)(30) & -0.585(15)(21)(38) \\ 
$(b\rightarrow 0)$ & & & &  \\ \hline
$32(4\pi)^2 L_{KK}^{I=1}(f_K)$
& 7.3(5)(8) & 6.8(3)(8) & 7.7(2)(8) & 7.4(3)(8)
\end{tabular}
\end{ruledtabular}
\end{table}

%
%
\begin{table}[t]
\caption{\label{tab:mKaKK_FINE} \textit{
The continuum limit of the scattering length at the physical point on the
fine MILC lattices, the
extracted counterterm that enters at NLO in $\chi$-PT, and the various
systematic uncertainties that have been identified beyond those associated with
fitting.
The correction factors and uncertainties are discussed in the caption of 
Table~\protect\ref{tab:mKaKK}.
}}
\begin{tabular}{c|c}
\hline\hline
Quantity & $m_l = 0.0062$ 
\\ \hline
$\Delta_{MA} \left(m_K a_{KK}^{I=1} \right)$ & -0.0048(15) \\
$\Delta_{NNLO}\left(m_K a_{KK}^{I=1} \right)$  & $\pm 0.013$ \\
$\Delta_{FV} \left(m_K a_{KK}^{I=1} \right)$ & $\pm 0.001$ \\
$\Delta_{m_{res}} \left(m_K a_{KK}^{I=1} \right)$ & $\pm 0.004$ \\
$\Delta_{range}\left(m_K a_{KK}^{I=1} \right)$ & $\pm 0.004$ \\
\hline
$m_{K^+} a_{K^+ K^+}$ & -0.387(28)(82)(14) \\ 
$(b\rightarrow 0)$ & \\ \hline
$32(4\pi)^2 L_{KK}^{I=1}(f_K)$ & 8.4(9)(2.6) 
\\ \hline\hline
\end{tabular}
\end{table}

%
%
\subsubsection{NNLO $\chi$-PT Corrections}

\noindent
The NNLO extrapolation formula for $m_K a_{KK}^{I=1}$ does not exist,
and therefore estimates of contributions from higher order in the chiral
expansion  are limited to  power-counting arguments.
A conservative estimate is provided by 
\begin{equation}
\Delta_{NNLO}\left(m_K a_{KK}^{I=1} \right) 
        = \pm \frac{2\pi m_K^6}{(4\pi f_K)^6} \left[ \ln \left(
            \frac{m_K^2}{f_K^2} \right) \right]^2
\ \ \ ,
\end{equation}
and the resulting uncertainties  are given in Table~\ref{tab:mKaKK}  and Table~\ref{tab:mKaKK_FINE}.

%
%
\subsubsection{Finite-Volume Effects in Mixed-Action $\chi$-PT}

\noindent
L\"{u}scher's relation between the two-particle energy levels in a
finite volume and their infinite-volume scattering parameters receive
exponential corrections which depend upon the lattice size and the
lightest particle in the spectrum, and generically scale as $e^{-m_\pi
L}$.  In Ref.~\cite{Bedaque:2006yi}, the exponential volume
corrections were determined for the $I=2\ \pi\pi$ system.  Using these
methods, one can also determine the exponential volume corrections for
the mixed-action $K^+ K^+$ system.  These exponentially-suppressed
volume corrections are formally sub-leading compared to the effective-range
corrections which have not been included, and 
provide an estimate of the finite-volume corrections. 
These terms are denoted as
\begin{equation}
        \Delta_{FV}\left(m_K a_{KK}^{I=1} \right) = \pm \left(
                m_K a_{KK}^{I=1} \Big|_{FV}
                -m_K a_{KK}^{I=1} \Big|_{\infty V} \right)\, .
\end{equation}
and are collected in Table~\ref{tab:mKaKK}  and Table~\ref{tab:mKaKK_FINE}.

%
%
\subsubsection{Residual Chiral Symmetry Breaking}

\noindent
The NLO mixed-action formula, eq.~\eqref{eq:mKaKKMA}, as well as the corrections of
Table~\ref{tab:mKaKK}  and Table~\ref{tab:mKaKK_FINE}, were derived assuming valence fermions with
perfect chiral symmetry.  
However, domain-wall fermions are
necessarily implemented  with a finite fifth-dimension  which induces 
residual chiral symmetry breaking.
The leading contributions from this residual chiral symmetry breaking
can be parameterized with a residual quark
mass~\cite{Shamir:1993zy,Furman:1994ky},
\begin{align}
        m_l^{dwf} &\rightarrow m_l^{dwf} + m_{res}\, ,
        \nonumber\\
        m_s^{dwf} &\rightarrow m_s^{dwf} + m_{res}\, .
\end{align}
However, by expressing the MA$\chi$-PT formula in terms of the
lattice-physical meson masses, the dominant contribution from these
$m_{res}$ terms are automatically included.  This leaves corrections at NLO
(assuming $m_{res} \sim m_q$ in the expansion), some of which have
undetermined coefficients.  Naive dimensional
analysis~\cite{Manohar:1983md} can be used to estimate the size of these terms,
\begin{equation}
        \Delta_{m_{res}} \left(m_K a_{KK}^{I=1} \right)
                =\pm \frac{8 \pi m_K^4}{(4\pi f_K)^4} \frac{m_{res}}{m_l}\, ,
\end{equation}
which are shown in Table~\ref{tab:mKaKK}  and Table~\ref{tab:mKaKK_FINE}.

%
%
\subsubsection{Range Corrections}

\noindent
When the spatial dimensions of the lattice are large compared to the range of
the interaction, and the scattering length is of natural size, 
as is the case for $K^+ K^+$ scattering at the quark masses used in this work, 
the effective range first enters at $\mathcal{O}(L^{-6})$ in the expansion of
the two-hadron energy in powers of $1/L$.
Therefore, neglecting the effective-range parameter introduces a $\sim 0.2\%
$ uncertainty in the extracted values of 
$m_{K^+} a_{K^+ K^+}$, assuming $r\sim 1/(2 m_\pi)$ for $m_\pi \ll m_K$. 
To be conservative, a $1\%$ systematic uncertainty due to the neglect of the
effective range 
is assigned to the scattering length determined  on each ensemble.

%
%
\subsection{Extrapolation to the Physical Point
\label{sec:mKaKKPred}}

%
%
\begin{figure}[!t]
\vskip0.5in
\center
\begin{tabular}{c}
\includegraphics[width=0.80\textwidth]{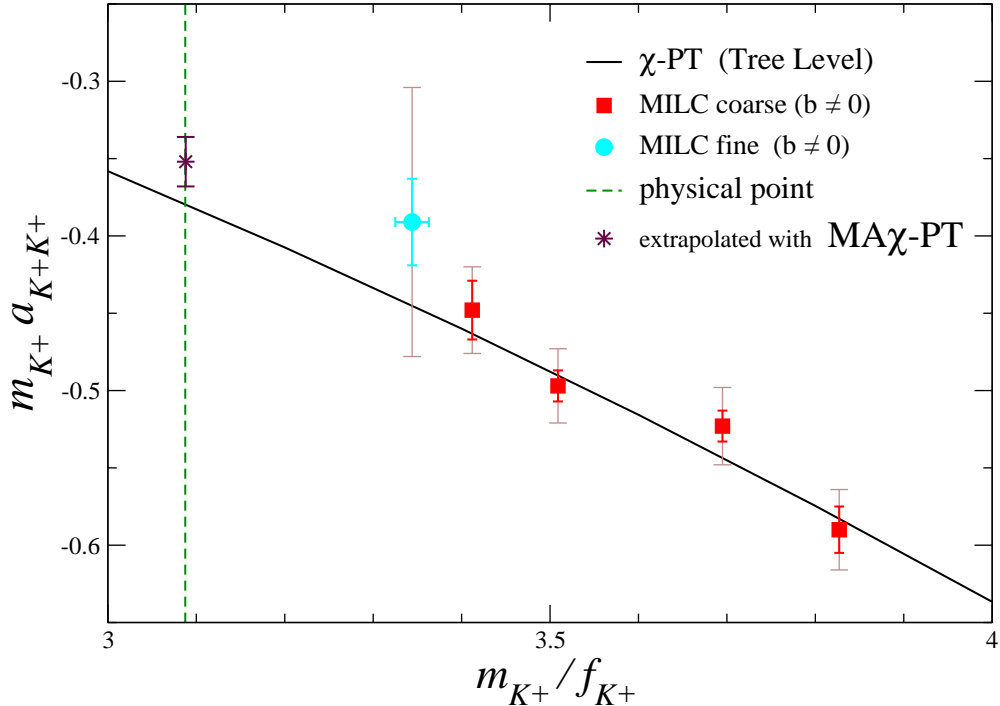} \\
\end{tabular}
\caption{\label{fig:mKaKKdataPhys} \textit{
$m_{K^+} a_{K^+ K^+}$ versus $m_{K^+}/f_{K^+}$.
The points with error-bars are the results of this lattice calculation (not
extrapolated to the continuum) on both the coarse and fine MILC lattices.
The solid curve corresponds to the tree-level prediction of $\chi$-PT, and the
point denoted by a star and its associated uncertainty is the value
extrapolated to the physical meson masses and to the continuum.
The smaller uncertainty associated with each point is statistical, while the
larger uncertainty is the statistical and  fitting systematic combined in quadrature.
}}
\end{figure}

\noindent
Calculations on the four coarse lattice ensembles yield pion and kaon
masses of approximately
$(m_\pi,m_K)\sim (290,580) , (350,595) , (490, 640)$ and $(590, 675)~{\rm
  MeV}$.
The chiral expansion will converge better for smaller meson masses, and 
one method to examine the convergence of the
chiral expansion is to selectively ``prune" the heaviest data
sets~\cite{Beane:2005rj,Beane:2006gj,Beane:2007xs}.
This is done by
first determining $L_{KK}^{I=1}(\mu = f_K)$ by fitting to all four
data points (fit A), then removing the heaviest point and fitting
(fit B) and finally removing the heaviest two points and fitting
(fit C).  The results of these fits are collected in Table~\ref{tab:L_GL}.
%
%
\begin{table}[t]
\caption{\label{tab:L_GL} \textit{
The results of  fitting three-flavor MA$\chi$-PT at NLO to the 
computed scattering lengths, as described in the text.
The values of $m_{K^+} a_{K^+ K^+}$ are those extrapolated to the physical
(isospin-symmetric) meson masses and to the continuum.
The first uncertainty is statistical and the second is 
systematic (as described in the text).}}
\begin{ruledtabular}
\begin{tabular}{c|ccc}
\ \ \ \ FIT\ \ \ \  & $32(4\pi)^2 L_{KK}^{I=1}(f_K)$ & $m_{K^+} a_{K^+ K^+}$ (extrapolated) & $\chi^2$/dof \\ \hline
A & 7.3(1)(4) & $-0.347 \pm 0.003 \pm 0.009$ & 0.22  \\
B & 7.3(2)(5) & $-0.347 \pm 0.004 \pm 0.011$ & 0.32  \\
C & 6.9(2)(6) & $-0.355 \pm 0.005 \pm 0.013$ & 0.14  \\ 
\end{tabular}
\end{ruledtabular}
\end{table}
The extracted values of $L_{KK}^{I=1}$ from each of the fits are consistent
with each other within the uncertainties.
In analogy with the comparison convention employed for $\pi^+\pi^+$,
the lattice data is extrapolated to the physical values of 
$m_{\pi^+}/f_{K^+} = 0.8731 \pm 0.0096$, $m_{K^+}/f_{K^+} =
3.088 \pm 0.018$ and $m_\eta / f_{K^+} = 3.425 \pm 0.0019$ assuming isospin
symmetry, and the absence of electromagnetism.
Taking the range of values of $L_{KK}^{I=1}$
spanned by these fits, we find
\begin{eqnarray}
          m_{K^+} a_{K^+ K^+} \ =\  -0.352 \pm 0.016
 \ \ \ \ , \ \ \ 
32(4\pi)^2 L_{KK}^{I=1}(\mu=f_K) & = & 7.1 \pm 0.7
\, ,
\end{eqnarray}
where the statistical and systematic errors have been combined in quadrature.
The results are shown in Fig.~\ref{fig:mKaKKdataPhys}.
It is somewhat surprising that the calculated scattering lengths
are consistent, within uncertainties, with tree-level $\chi$-PT.
This was also found to be the case for $\pi^+\pi^+$ scattering even at large
pion masses.

%
%
\subsection{Comparing $K^+ K^+$ scattering with $\pi^+ \pi^+$ scattering}

\noindent
A comparison between the  lattice calculations of
$\pi^+\pi^+$~\cite{Beane:2007xs}  
and $K^+ K^+$ scattering lengths 
allows for a study of flavor-$SU(3)$ breaking in the scattering amplitude
due to terms that are beyond NLO in $\chi$-PT.
The linear combination of Gasser-Leutwyler coefficients contributing
to the $I=2\ \pi\pi$ scattering length at NLO is the same as the
combination contributing to the $I=1\ KK$ scattering
length~\cite{Chen:2006wf}:
\begin{equation}
        L_{KK}^{I=1}(\mu) =     L_{\pi\pi}^{I=2}(\mu)\, .
\end{equation}
To compare extractions of these counterterms, the scales at which they are evaluated
must be the same, and   the scale-dependence  of $L_{KK}^{I=1}(\mu)$ is
\begin{equation}
        32(4\pi)^2 L_{KK}^{I=1}(\mu) = 
                32(4\pi)^2 L_{KK}^{I=1}(\mu_0) -\frac{28}{9} \ln \left(
                  \frac{\mu^2}{\mu_0^2} \right)
\, .
\end{equation}
\begin{figure}[!t]
\vskip0.5in
\center
\begin{tabular}{c}
\includegraphics[width=0.6\textwidth]{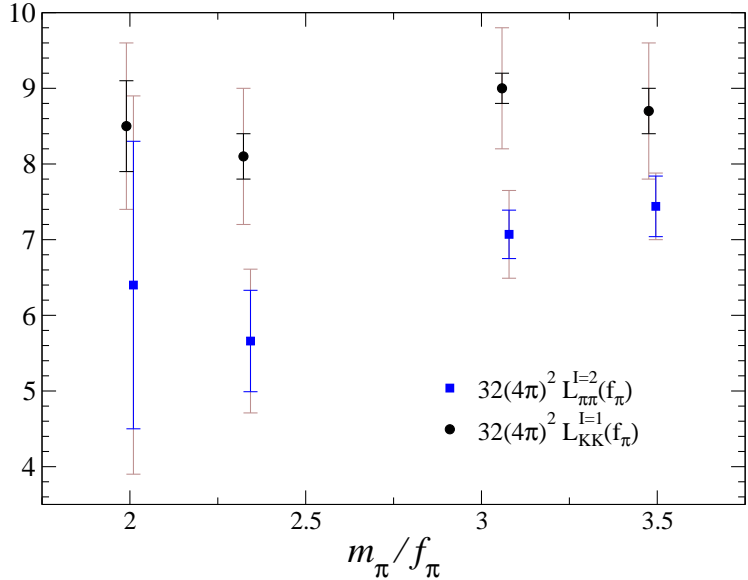} \\
\end{tabular}
\caption{\label{fig:SU3compare} \textit{ $32(4\pi)^2
L_{KK}^{I=1}(f_\pi)$ (circles) and $32(4\pi)^2
L_{\pi\pi}^{I=2}(f_\pi)$ (squares) versus $m_{\pi}/f_{\pi}$. The sets
have been slightly displaced horizontally for convenience of viewing. The smaller
uncertainty associated with each point is statistical, while the
larger uncertainty is the statistical and fitting systematic combined
in quadrature.  }}
\end{figure}
The counterterms extracted from the  mixed-action lattice calculations 
are shown in fig.~\ref{fig:SU3compare} as a function of
$m_\pi/f_\pi$.
It is clear that while there appears to be a
difference between $L_{KK}^{I=1}(f_\pi)$ and $L_{\pi\pi}^{I=2}(f_\pi)$, 
more precise calculations of both scattering lengths, particularly at the
lightest pion masses, are required for further exploration of higher order
terms in the chiral expansion.

%
%
\section{Discussion \label{sec:Conclude}}

\noindent
We have presented results of a lattice QCD calculation of
the $K^+ K^+$ scattering length performed with domain-wall
valence quarks on asqtad-improved MILC configurations with 2+1
dynamical staggered quarks.  The calculations were performed on the coarse 
MILC lattices with a 
lattice spacing of $b\sim 0.125~{\rm fm}$ (with a preliminary calculation on
one ensemble of the fine MILC lattices with $b\sim 0.09~{\rm fm}$)
and at a single lattice spatial size of $L\sim 2.5~{\rm fm}$.
One-loop
MA$\chi$-PT with three flavors of light quarks was used to perform the
chiral and continuum extrapolations.  
Our prediction for the physical value of the  $K^+ K^+$  scattering length is 
$m_{K^+}\  a_{K^+ K^+} = -0.352 \pm 0.016$, and we emphasize
once again that this result rests on the assumption that the
fourth-root trick recovers the correct continuum limit of QCD. 
Deviations from Weinberg's tree-level prediction are found to be
surprisingly small, consistent with the lattice calculations of the $\pi^+\pi^+$ 
scattering length at heavier pion masses.

\section{Acknowledgments}

\noindent We thank R.~Edwards and B.~Joo for help with the QDP++/Chroma
programming environment~\cite{Edwards:2004sx} with which the
calculations discussed here were performed.  The computations for this work
were performed at Jefferson Lab, Fermilab, Lawrence Livermore National
Laboratory, National Center for Supercomputing Applications, and
Centro Nacional de Supercomputaci\'on (Barcelona, Spain).  We are
indebted to the MILC and the LHP collaborations for use of their
configurations and propagators, respectively.  The work of MJS was
supported in part by the U.S.~Dept.~of Energy under Grant
No.~DE-FG03-97ER4014. The work of KO was supported in part by the
U.S.~Dept.~of Energy contract No.~DE-AC05-06OR23177 (JSA) and 
by the Jeffress Memorial Trust,
grant J-813.  The work of AWL was supported in part by the
U.S.~Dept.~of Energy grant No.~DE-FG02-93ER-40762.  The work of SRB
and AT was supported in part by the National Science Foundation 
CAREER grant No. PHY-0645570. Part of this work was performed under the
auspices of the US DOE by the University of California, Lawrence
Livermore National Laboratory under Contract No. W-7405-Eng-48.  The
work of AP was partly supported by the EU contract FLAVIAnet
MRTN-CT-2006-035482, by the contract FIS2005-03142 from MEC (Spain)
and FEDER and by the Generalitat de Catalunya contract 2005SGR-00343.

%
%

\end{document}